\documentstyle[12pt,epsfig,axodraw]{article}
\let\section=\subsection     \let\subsection=\subsubsection           
                      
\newcommand{\M}{{\cal M}}
\newcommand{\MeV}{{\rm MeV}}
\newcommand{\GeV}{{\rm GeV}}

\newcommand{\fm}{{\rm fm}}
\renewcommand{\Im}{{\rm Im}}

\newcommand{\gsim}{$\raisebox{-0.8ex} {$\stackrel{\textstyle >}{\sim}$}$}

\begin{document}
\begin{center}
   {\large \bf VECTOR MESONS IN MEDIUM} \footnote{Work supported in part by GSI and BMBF}\\[5mm]
   F. Klingl and W. Weise  \\[5mm]
   {\small \it  Physik-Department\\ Technische Universit\"at
M\"unchen \\ D-85747 Garching, Germany\\[8mm]}
\end{center}

\begin{abstract}\noindent
  Based on an effective Lagrangian which combines chiral SU(3) dynamics with
  vector meson dominance, we have developed a model for the forward vector
  meson-nucleon scattering amplitudes. We use this as an input to calculate the
  low energy part of the current-current correlation function in nuclear
  matter. For the isovector channel we find a significant enhancement of the
  in-medium spectral density below the $\rho$ resonance while the $\rho$ meson
  mass itself changes only slightly. The situation is different in the
  isoscalar channel, where the mass and peak position of the $\omega$ meson
  move downward while its width increases moderately. For the $\phi$ meson we
  find almost no mass shift, but the width of the peak increases significantly. We use
  these spectra as ``left hand side'' of in medium QCD sum rules and observe a
  remarkable degree of consistency with the operator product expansion ``right
  hand''side in
  all three channels.  We point out, however, that these results cannot simply
  be interpreted, as commonly done, in terms of a universal rescaling of vector
  meson masses in matter. We also compare the resulting thermal dilepton
  rates of our model with CERES data. We find satisfactory agreement
  but we point out that the dilepton rates from completely uncorrelated
  quark-antiquark pairs also fit the data quite well. Considering the strongly attractive potential
  for the $\omega$ together with its comparably small width, we discuss the
  possible 
  formation of bound $\omega$ states in nuclei. Using a Green's function approach we
  investigate the possibility of detecting these bound states in the $A(d, ^3\!He)
  $ pick-up reaction.
\end{abstract}

\section{Introduction}
It is well known that the chiral symmetry of QCD in spontaneously broken. This
leads to the octet of light pseudoscalar mesons which are the Goldstone bosons
of this broken symmetry, and to a mass splitting between chiral partners of
all hadrons. An example is the vector meson whose chiral partner, the axial
vector meson, is twice as heavy. There are, however, strong hints from the
quark condensate in medium \cite{1}, QCD lattice simulations \cite{2,3} or
schematic chiral models such as the Nambu-Jona-Lasinio model \cite{4}, that
this symmetry is expected to be
restored at high temperatures and high densities. In order to find some signals
for this chiral restoration it was proposed to study dileptons emitted in heavy
ion collisions since they can carry information of the inner hot and dense zone
of such collisions. Such experiments with high statistics and resolution are
planned by the HADES collaboration at GSI. For high temperature there are
already data from the CERES \cite{5} and HELIOS \cite{6} collaborations at
CERN which observe strongly enhanced dilepton yields below the rho meson
resonance. This was often interpreted as an indication of dropping vector meson
masses which in addition should be a signal for chiral restoration. The
Brown-Rho scaling scenario \cite{7} and considerations based on bag models \cite{8}
support this idea and find a strong reduction of the vector mesons masses. Also
the QCD sum rule analysis \cite{9}, which however only used a delta
function caricature of the true spectrum, seemed to confirm these results.
From the theoretical point of view chiral restoration, however, does not demand
a drastic reduction of the vector mesons masses in medium. Hadronic calculations
\cite{10,11,12} suggest only marginal changes of the in-medium rho meson mass,
but a strongly increased decay width instead. It was shown \cite{12,13} that
this is also in agreement with QCD sum rule analysis if one replaces the
delta function by the full spectrum. Taking also p-wave interactions \cite{14}
into account this leads to an agreement with the CERES dilepton data.

In a thermal model the dilepton yields of heavy ion collisions are governed
by the thermal dilepton rates given by
\begin{equation}
  \label{eq1}
   \frac{dR}{d^4x \, d^4q} = -\frac{ \alpha^2}{\pi^3 q^2} 
  \Im \, \Pi (q,\, \rho,\, T) \, f_B(q,T), 
\end{equation}
with $\alpha=e^2/4 \pi=1/137$.
We denote the four-momentum of the dilepton pair by $q$, the temperature and the
density of the system with $T$ and $\rho$. The space-time volume element which radiates
the dilepton pair is $d^4x$. Apart from the trivial Boltzmann factor $f_B$, only the imaginary part of the current-current
correlation function $\Pi (q,\, \rho,\, T)$ , which we will investigate in this
talk, determines directly the
dilepton rates. 

In section 2 we review the properties of the current-current correlation function in
the vacuum. The corresponding correlation functions in baryonic matter will be developed
in section 3. To leading order in the density this
requires a detailed calculation of the vector meson-nucleon scattering
amplitudes. The in-medium spectra
and their comparison with QCD sum rules will then be presented in chapter 4. In section 5 we
compare the resulting dilepton rates with the CERES data. We also show the
result from the correlation function which originates from uncorrelated
quark-antiquark pairs. We will see that both scenarios agree with present data. We
therefore suggest to study vector mesons in additional experiments under more
moderate conditions, such as the $A(d, ^3\!He)$
pick-up reaction, to gain further informations. 
\section{Vector mesons in the vacuum}

Before we start looking into the in-medium properties of neutral vector mesons
it is useful to give a brief reminder of their vacuum properties. Since the vector mesons are
not stable they can only be seen as resonances in the current-current (CC)
correlation function,
\begin{equation}
  \Pi_{\mu\nu}(q)=i\int d^4x \: e^{iq\cdot x}\langle
  0|{\cal T}j_{\mu}(x) j_{\nu}(0)|0\rangle  ,
 \label{2.1}
\end{equation} 
where ${\cal T}$ denotes the time-ordered product and $j_{\mu}$ is the
electromagnetic current. It can be decomposed as
\begin{equation}
  \label{2.3}
  j_{\mu}= j_\mu^\rho+j_\mu^\omega+j_\mu^\phi
\end{equation}
into vector currents specified by their quark
content:
\begin{eqnarray}
  j^\rho_{\mu}&=&\frac{1}{2}(\bar{u}\gamma_{\mu}u-\bar{d}\gamma_{\mu}d),\\
  j^\omega_{\mu}&=&\frac{1}{6}(\bar{u}\gamma_{\mu}u+\bar{d}\gamma_{\mu}d),\\
  j^\phi_{\mu}&=&-\frac{1}{3}(\bar{s}\gamma_{\mu}s).
\label{2.2}
\end{eqnarray}
Current conservation implies a transverse tensor structure
\begin{equation}
  \Pi_{\mu\nu}(q)=\left(g_{\mu\nu}-\frac{q_{\mu}q_{\nu}}{q^2}\right)\Pi(q^2),
\label{2.4}
\end{equation}
with the scalar CC correlation function
\begin{equation}
  \Pi (q^2)=\frac{1}{3} g^{\mu\nu}\Pi_{\mu\nu}(q).
\label{2.5}
\end{equation}
\begin{figure}[h]
\unitlength0.7071mm
\begin{picture}(208,90)
\put(0,0){\makebox{\epsfig{file=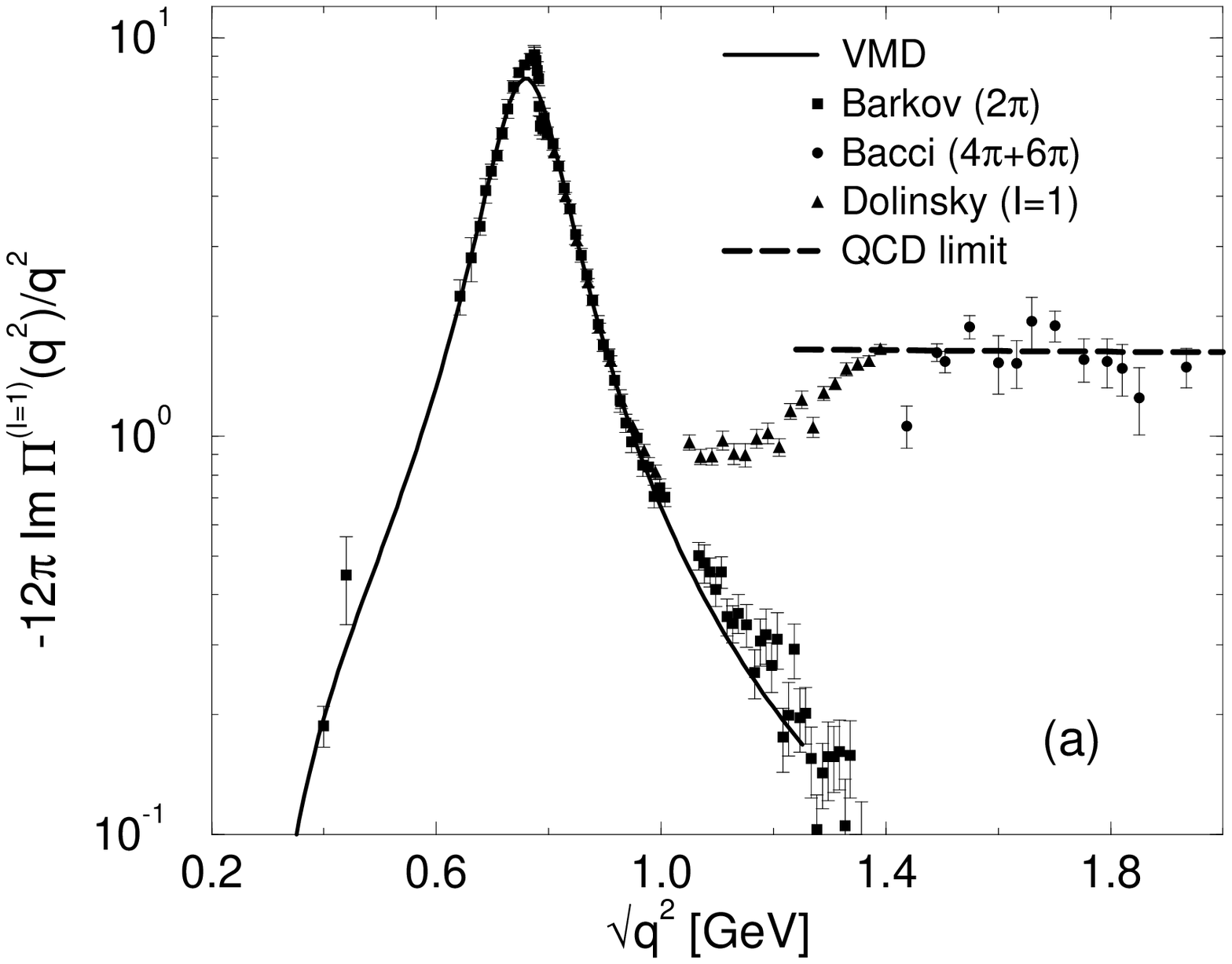,width=75mm}}}
\put(100,0){\makebox{\epsfig{file=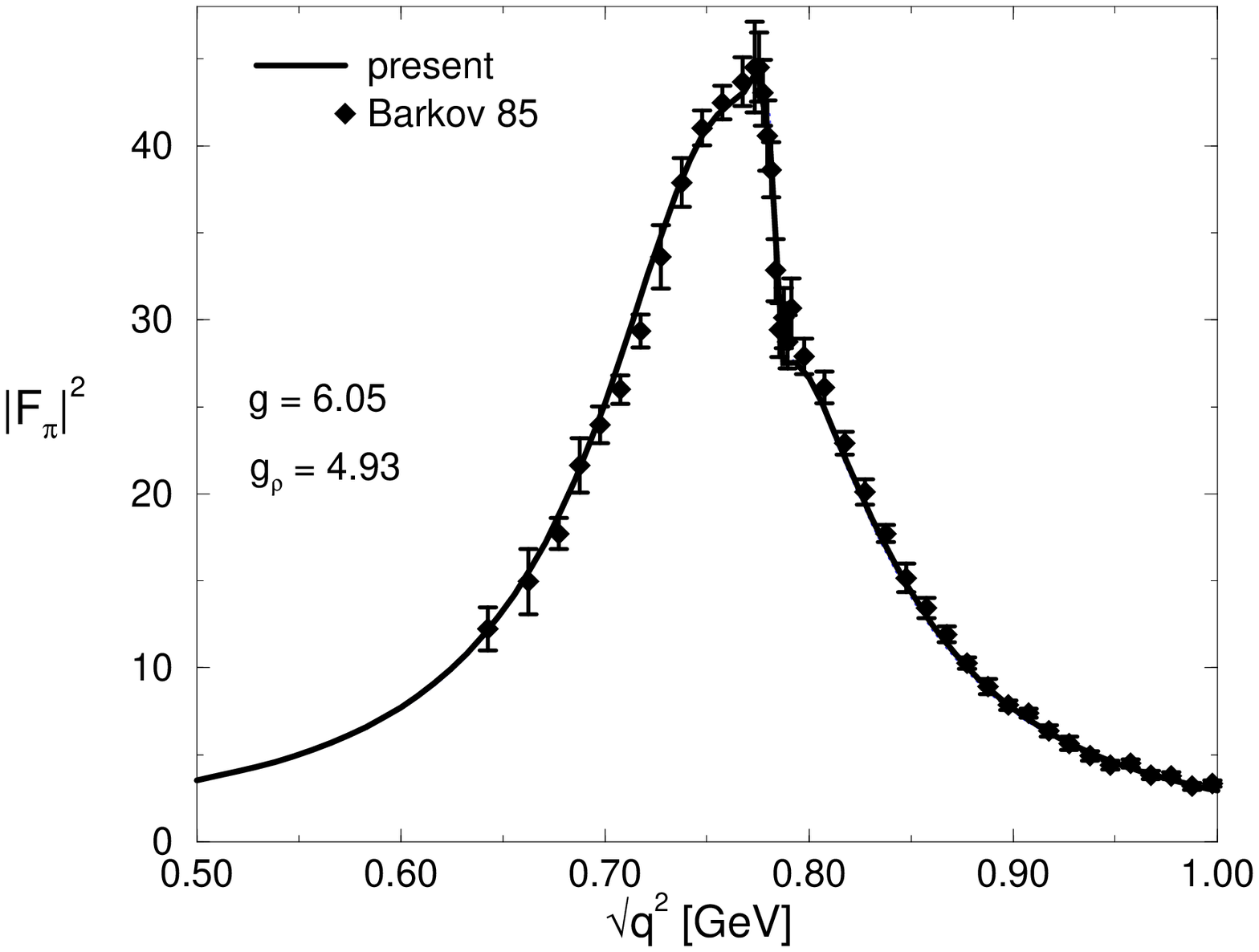,width=75mm}}}

\put(30,85){\makebox{a: Spectrum}}
\put(130,85){\makebox{b: Pion form factor}}
\end{picture}
\vspace*{-0.3cm}
\caption{ Spectrum of the isovector current-current correlation function in the vacuum and
the pion form factor . The solid lines shows result from our vector meson
dominance calculation \protect \cite{15}. }
\vspace*{-0.2cm}
\end{figure}

The imaginary part of the correlation function is proportional to the cross
section for $e^+ e^- \to$ hadrons: 
\begin{equation}
 R(s)=\frac{\sigma (e^+ e^-\to \rm hadrons )}{\sigma (e^+
  e^-\to \mu^+ \mu^-)}=-\frac{12 \pi}{s} \Im \Pi(s)
\label{2.6}
\end{equation}
where $\sigma (e^+ e^-\to \mu^+ \mu^-)= 4 \pi \alpha^2/3 s$ with $\alpha=e^2/4
\pi=1/137$. The vector mesons can be distinguished by looking at different
hadronic channels with corresponding flavour (isospin) quantum numbers.
We show for example the isovector current describing the $\rho$
meson. G-parity demands that it decays only into even numbers of pions (see
data in fig.1a). We observe two different energy regions: the low energy region of the CC correlation
function is very well described by a hadronic language through vector meson dominance (VMD). The extended VMD model \cite{15} gives
\begin{equation}
  \label{2.7}
  \Im \Pi^{(I=1)} (q^2)= \frac{\Im \Pi_\rho^{\rm vac} (q^2)}{g^2} \left| F_\pi(q^2) \right|^2,
\end{equation}
which is shown by the solid line in Fig 1a. It is determined by the imaginary
part of the vacuum $\rho$-meson self energy $\Pi^{\rm
  vac}_\rho$ \cite{15}, which comes from the decay of the $\rho$
into two pions, and the pion form factor
\begin{equation}
  \label{2.9}
   F_\pi (q^2) = \frac{\left(1- \frac{g}{\stackrel{ \rm o }{g}_\rho}\right)q^2-\stackrel{ \rm o }{m}_\rho^2}{q^2-\stackrel{ \rm o }{m}_\rho^2-\Pi_\rho(q^2)}\, ,
\end{equation}
shown in fig 1b explicitly.
Here $\stackrel{ \rm o }{m}_\rho$ is the bare $\rho$ meson mass and $g$
($\stackrel{ \rm o }{g}_\rho$) is the coupling of the $\rho$ meson to the pion
(photon).  The ratio of $g$ to $g_\rho$ is identical to unity in case of
``complete'' VMD in which all of the hadronic electromagnetic interaction is
transmitted through vector mesons. Note in passing that the denominator of the
pion form factor is the full $\rho$ meson propagator, which originates from the
intermediate $\rho$-meson as depicted in fig.2 and gives rise to the peak in the pion
form factor at the rho meson mass. Isospin violating processes are small but
visible as $\rho \omega$ mixing corrections in the peak of the pion formfactor
(see Fig 1b). 

\begin{figure}[h]
\unitlength0.7071mm
\begin{picture}(200,50)
\put(0,0){\makebox{\epsfig{file=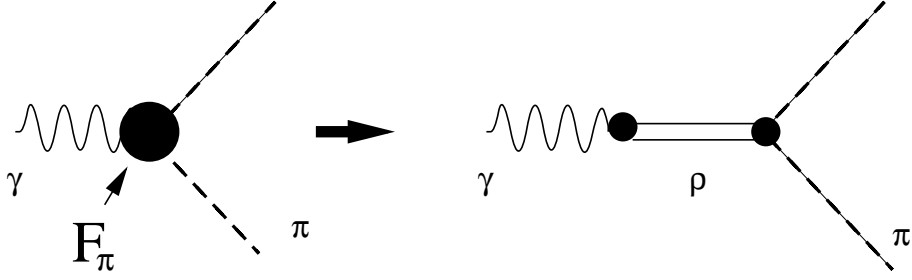,width=120mm}}}
\end{picture}
\caption{ Diagrammatic resolution of the pion form factor}
\end{figure}

In the high energy region on the other hand the measured correlation
function approaches the asymptotic plateau predicted by perturbative QCD:
\begin{equation}
 -\frac{12 \pi}{q^2} \Im \Pi(q^2) =\frac{3}{2} \left( 1+\frac{\alpha_S}{\pi} \right) \Theta (q^2- s_0)
\label{2.10}
\end{equation}
The threshold $s_0$ lies at about 1.2 $\GeV^2$. This continuum stems from a
picture where the photon couples directly to a quark-antiquark pair, which
travels freely over a short distance and then hadronizes into the multi pion continuum. Note that a system of
uncorrelated quark and gluons ("Quark-Gluon-Plasma") would also lead to such
a plateau in the spectrum. The threshold would then start already at twice
the current quark mass. 
\section{Current correlation functions in baryonic matter}

The CC-correlation function in medium at temperature $T=0$ is defined as
\begin{eqnarray}
  \label{3.1}
  \Pi_{\mu\nu}(\omega ,\vec{q};\rho)\: \: = \:\:  i\int\limits^{+\infty}_{-\infty}dt\int
  d^3 x  \: e^{i\omega t-i\vec{q}\cdot\vec{x}} \hspace{2cm}\\ \nonumber \hspace{3.cm}
*\,  \left<\rho\left|{\cal T}j_{\mu}(t,\vec{x})j_{\nu}(0)
  \right|\rho\right>,
\end{eqnarray}
where we have replaced the vacuum by $\big| \rho \rangle$, the ground state of infinite, isotropic and isospin
symmetric nuclear matter with density $\rho$. We assume matter as a whole to be
at rest. This specifies the
Lorentz frame that we will use in the following. For $\vec{q}=0$, the case where the vector meson is at rest, only the
transverse tensor structure survives and $\Pi^{00}=\Pi^{0j}=\Pi^{i0}=0$. We write
\begin{equation}
  \label{3.2}
  \Pi_{ij} (\omega ,\vec{q}=0;\rho)= -\delta_{ij}  \Pi (\omega,\vec{q}=0;\rho).
\end{equation}

\begin{figure}[h]
\vspace*{-4cm}
\begin{center}
\begin{picture}(400,300)(0,0)
\SetWidth{1}    
\ArrowLine(10,155)(30,155)
\ArrowLine(30,155)(70,155)
\ArrowLine(70,155)(90,155)
\DashCArc(50,155)(20,0,180){6}
\Photon(10,175)(30,155){3}{3}
\Photon(90,175)(70,155){3}{3}
\Text(50,145)[]{(a)}
\Text(50,185)[]{$\pi$,K}
\Text(50,165)[]{B}
\Vertex(30,155){2}
\Vertex(70,155){2}

\ArrowLine(110,155)(130,155)
\ArrowLine(130,155)(170,155)
\ArrowLine(170,155)(190,155)
\DashCArc(150,155)(20,0,180){6}
\Photon(116,189)(136,169){3}{3}
\Photon(164,169)(184,189){3}{3}
\Text(150,145)[]{(b)}
\Vertex(130,155){2}
\Vertex(170,155){2}
\Vertex(136,169){2}
\Vertex(164,169){2}

\ArrowLine(210,155)(230,155)
\ArrowLine(230,155)(270,155)
\ArrowLine(270,155)(290,155)
\DashCArc(250,155)(20,0,180){6}
\Photon(210,175)(230,155){3}{3}
\Photon(264,169)(284,189){3}{3}
\Text(250,145)[]{(c)}
\Vertex(230,155){2}
\Vertex(270,155){2}
\Vertex(264,169){2}

\ArrowLine(310,155)(330,155)
\ArrowLine(330,155)(370,155)
\ArrowLine(370,155)(390,155)
\DashCArc(350,155)(20,0,180){6}
\Photon(330,195)(350,175){3}{3}
\Photon(350,175)(370,195){3}{3}
\Text(350,145)[]{(d)}
\Vertex(330,155){2}
\Vertex(370,155){2}
\Vertex(350,175){2}

\ArrowLine(10,15)(30,15)
\ArrowLine(30,15)(70,15)
\ArrowLine(70,15)(90,15)
\Line(28,15)(28,55)
\Line(32,15)(32,55)
\Line(68,15)(68,55)
\Line(72,15)(72,55)
\DashLine(30,55)(70,55){4}
\Photon(10,75)(30,55){3}{3}
\Photon(70,55)(90,75){3}{3}
\Text(50,5)[]{(e)}
\Text(20,40)[]{$\rho$}
\Text(80,40)[]{$\rho$}
\Text(50,65)[]{$\pi$}
\Vertex(30,15){2}
\Vertex(70,15){2}
\Vertex(30,55){2}
\Vertex(70,55){2}

\ArrowLine(110,15)(130,15)
\ArrowLine(130,15)(170,15)
\ArrowLine(170,15)(190,15)
\Line(130,53)(170,53)
\Line(130,57)(170,57)
\DashLine(130,15)(130,55){4}
\DashLine(170,15)(170,55){4}
\Photon(110,75)(130,55){3}{3}
\Photon(170,55)(190,75){3}{3}
\Text(150,5)[]{(f)}
\Text(120,40)[]{$\pi$}
\Text(180,40)[]{$\pi$}
\Text(150,65)[]{$\rho$}
\Vertex(130,15){2}
\Vertex(170,15){2}
\Vertex(130,55){2}
\Vertex(170,55){2}

\ArrowLine(210,15)(230,15)
\ArrowLine(230,15)(270,15)
\ArrowLine(270,15)(290,15)
\Line(228,15)(228,55)
\Line(232,15)(232,55)
\Line(268,15)(268,55)
\Line(272,15)(272,55)
\DashLine(230,55)(270,55){4}
\DashCArc(250,15)(20,0,180){6}
\Photon(210,75)(230,55){3}{3}
\Photon(270,55)(290,75){3}{3}
\Text(250,5)[]{(g)}
\Text(220,40)[]{$\rho$}
\Text(280,40)[]{$\rho$}
\Text(250,65)[]{$\pi$}
\Vertex(230,15){2}
\Vertex(270,15){2}
\Vertex(230,55){2}
\Vertex(270,55){2}

\Photon(310,75)(330,55){3}{3}
\DashLine(330,55)(370,55){3}
\Photon(370,55)(390,75){3}{3}
\DashLine(330,55)(350,15){3}
\DashLine(350,15)(370,55){3}
\ArrowLine(325,15)(350,15)
\ArrowLine(350,15)(375,15)
\Text(346,5)[]{\large (h)}
\Text(345,65)[]{ K}
\Text(325,37)[]{ K}
\Text(312,59)[]{ $\phi$}

\end{picture}
\end{center}
\vspace*{-0.3cm}
\caption{Dominant diagrams for the Compton amplitude $T(\omega)$ and the vector
meson-nucleon amplitudes $T_{VN}(\omega)$. In order to get from $T$ to $T_{VN}$,
replace the (wavy) photon line by the relevant vector mesons. Diagrams (e) and
(f) operate, in particular, in the $\omega$ meson channel.}
\end{figure}
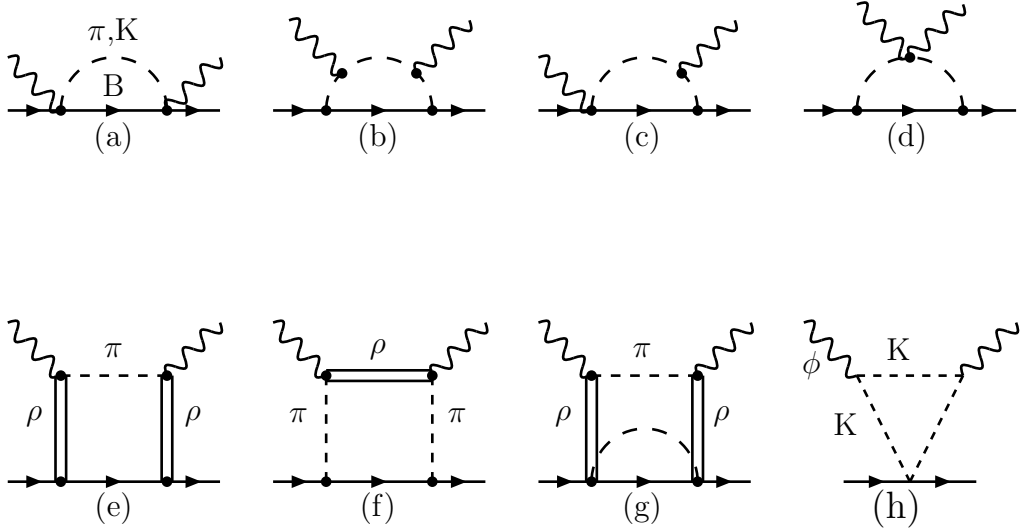

As a first step we apply the low density theorem \cite{16} to the self energy of the vector mesons:
\begin{equation}
\label{3.4}
  \Pi_V (\omega ,\vec{q}=0;\rho)= \Pi_V^{\rm vac}(\omega^2)-\rho \, T_{VN} (
  \omega)+... \, ,
\end{equation}
where $T_{VN}$ is the vector meson-nucleon scattering amplitude taken at
$\vec{q}=0$. We can extend eq.(\ref{2.7}) to the in-medium correlation function
in terms of VMD and write \cite{12}
\begin{eqnarray}
  \label{3.5}
   \Im \Pi(\omega,\vec{q}=0;\rho)\: \: = \: \frac{1}{g_V^2} \Im \left(\Pi^{\rm vac}_V(\omega^2)-\rho
   T_{V N}(\omega)\right)\\ \nonumber \hspace{2.0cm} * \;\left|\frac{(1-a_V) \,\omega^2-\stackrel{ \rm o
  }{m}_V^2}{\omega^2-\stackrel{ \rm o }{m}_V^2-\left(\Pi^{\rm
  vac}_V(\omega^2) -\rho \, T_{VN}(\omega)\right)}\right|^2.
\end{eqnarray}

\begin{figure}[h]
\vspace*{-0.2cm}
\unitlength0.7071mm
\begin{picture}(208,180)
\put(50,90){\makebox{\epsfig{file=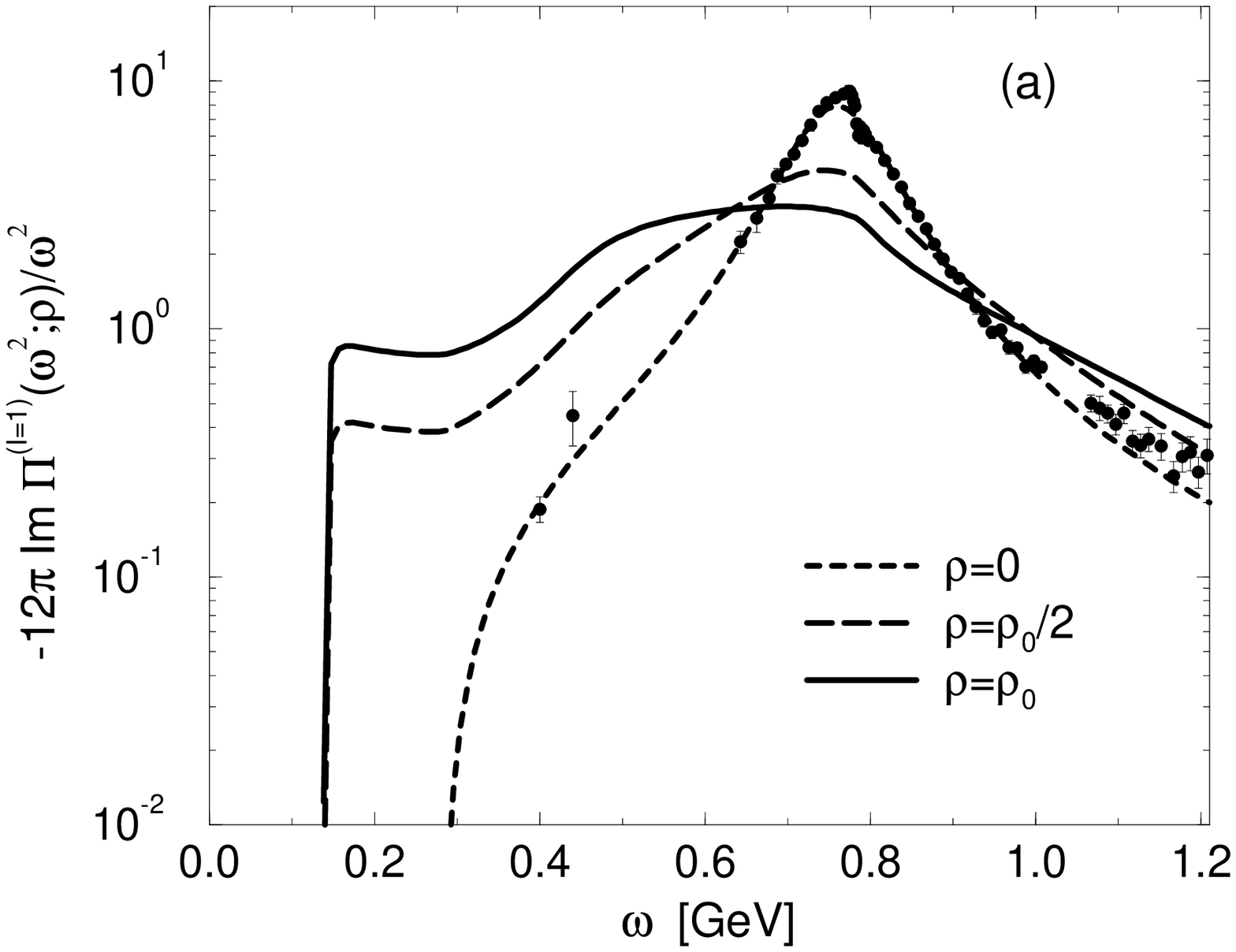,width=75mm}}}
\put(0,0){\makebox{\epsfig{file=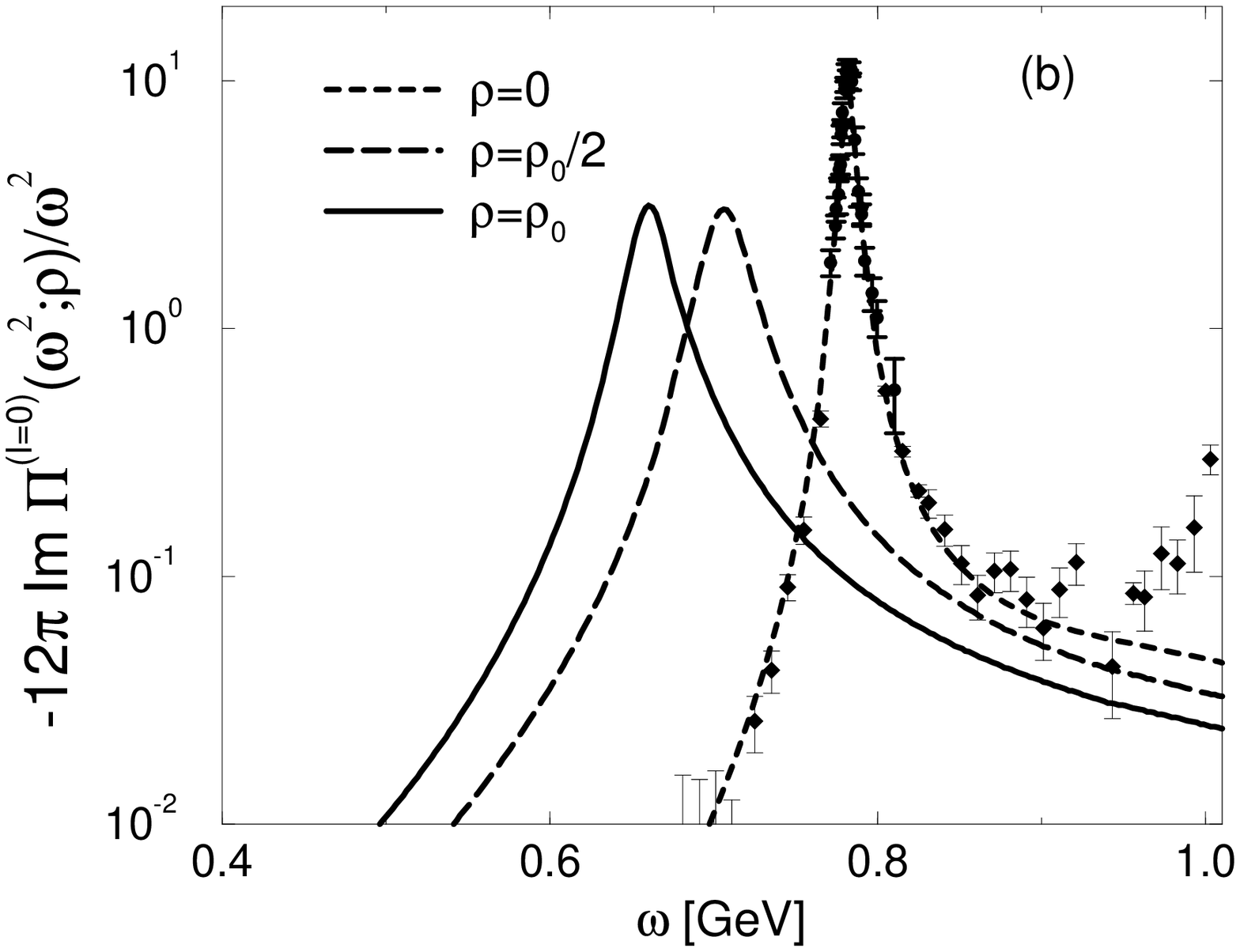,width=75mm}}}
\put(100,0){\makebox{\epsfig{file=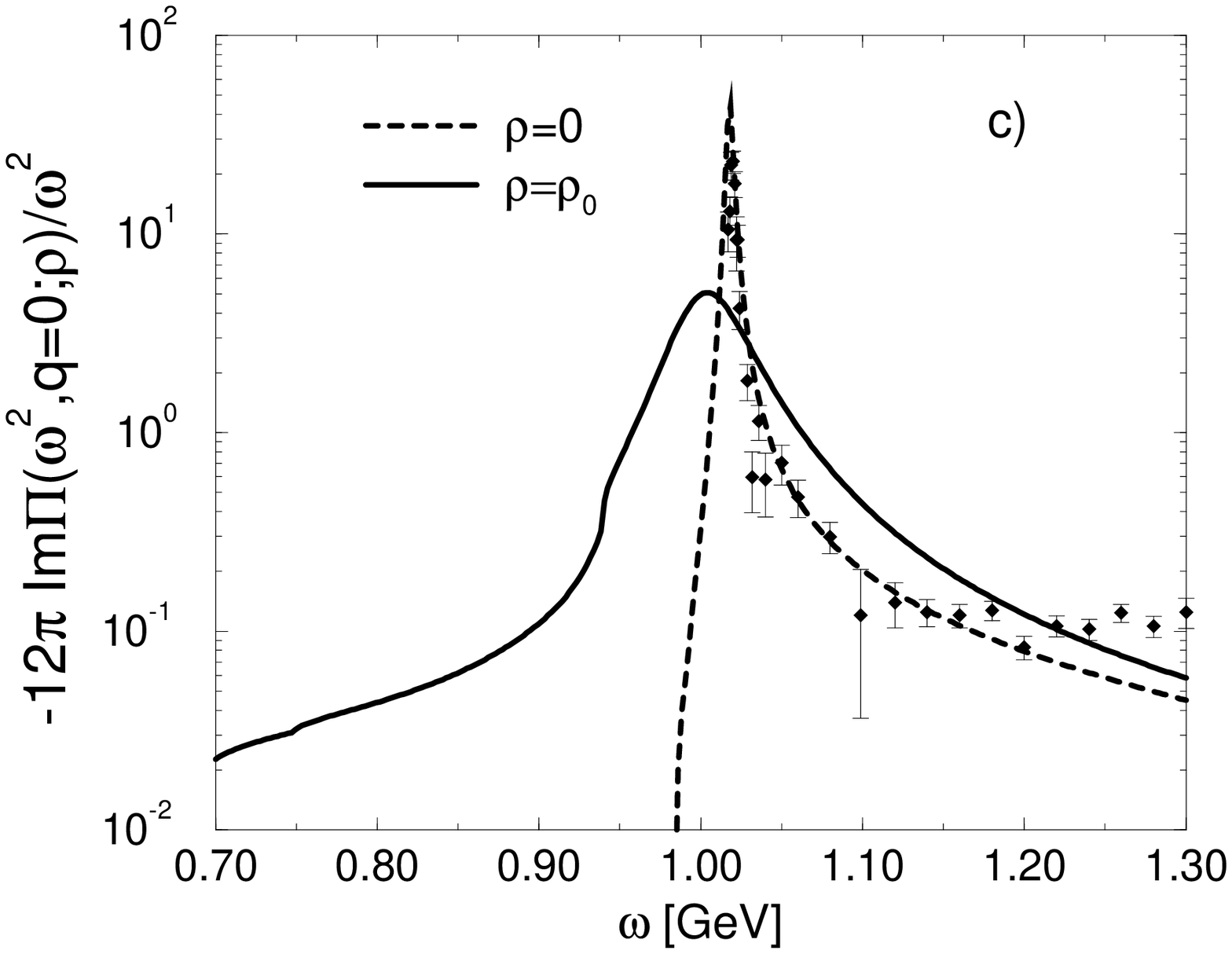,width=75mm}}}

\put(95,175){\makebox{a: $\rho$ meson}}
\put(45,85){\makebox{b: $\omega$ meson}}
\put(145,85){\makebox{c: $\phi$ meson}}
\end{picture}
\caption{Calculated spectra of current-current correlation functions. The dashed
lines show the vacuum spectra in the $\rho$, $\omega$ and $\phi$ channels
normalized such that they can be compared directly with the corresponding
$e^+e^- \to$ hadrons data. The long dashed and solid lines show the spectral functions in
nuclear matter at densities $\rho_0/2$ and $\rho_0=0.17 \; \fm^{-3}$, as
discussed in section 3.}
\vspace*{-0.2cm}
\end{figure}

We wrote down here the general form for all three channels replacing $g$ by
$g_V$ and denoting the ratio by $a_V$. The vacuum vector meson self energy was
replaced by eq.(\ref{3.4}) which includes the complete set of medium
modifications.  In order to determine the in-medium correlation function we are
left with the task to calculate the vector meson-nucleon scattering amplitudes.
We use an effective Lagrangian which combines chiral SU(3) dynamics with VMD.
This Lagrangian has been applied successfully in the vacuum \cite{15} and has
been extended to incorporate meson-baryon interactions. For the baryons we
include the complete baryon octet and decuplet (nucleons,
hyperons, $\Delta$'s, etc.).

The most important processes contributing to the scattering amplitudes are
shown in fig.3 for the $\rho$, $\omega$ and $\phi$ meson respectively.  For the
$\rho$ ($\phi$) meson we draw the diagrams which survive in the limit of large
baryon mass (fig. 3a-d). The last one (fig. 3d) only contributes to the real
part of the $\rho N$ ($\phi N$) scattering amplitude. While for the
$\rho$-meson the $\pi N$ and $\pi \Delta$ loops govern the scattering
amplitude, $K\Sigma$, $K\Lambda$ and $K\Sigma(1385)$ loops dominate for the
$\phi$ meson. We also include s-wave scattering of the kaons (fig. 3h)
\cite{17}. For the $\omega N $ scattering amplitude we only show the dominant
contributions to the imaginary part (figs. 3e, f, g). We evaluate the imaginary
parts of the amplitudes by cutting the diagrams in all possible ways. The real
parts are then determined by using a once subtracted dispersion relation, with
the subtraction constant fixed by the Thomson limit. Evaluating those diagrams
and using eq.(\ref{3.5}) we plot the spectra of the correlation functions for
various densities as shown by the curves in fig. 4a, b and c. We observe
important differences between the various channels. The $\rho$ meson mass
decreases only slightly while the width increases very strongly. This causes
the peak to shift downwards and broaden. We also see strong threshold
contributions starting at the pion mass. The $\rho$-meson therefore dissolves
in nuclear matter; it is not a good quasi-particle any longer. On the other hand
the mass of the $\omega$ meson decreases significantly. Its width also
increases but not as strongly as for the $\rho$ meson and so it can still be
regarded as a good-quasi particle. For the $\phi$ meson there is almost no
change in the peak position while the width increases.

\section{Comparison with QCD sum rules}
Up to now we only looked at the time like region of the correlation function
where it has an imaginary part and is therefore experimentally accessible. Now we
want to concentrate on the spacelike region. For momentum $|q^2| \gg 1 \; \GeV$
the correlation function is governed by
perturbative QCD which and is well under control. The interesting region is the
space like region at around $Q^2=-q^2 \simeq 1 \; \GeV$ where non-perturbative
effect start to become important. The basic idea of QCD sum rules \cite{18} is to compare
two ways of determining $\Pi(q^2)$ in this region.
One way is to use a twice
subtracted dispersion relation for each one of the channels $V=\rho$,
$\omega$, $\phi$:
\begin{equation}
  \Pi(q^2)=\Pi (0)+c \, q^2+\frac{q^4}{\pi}\int ds\frac{\Im\Pi (s)}{s^2(s-q^2
  -i\epsilon)}, 
\label{2.12}
\end{equation}
where one uses the measured or calculated spectrum $\Im \Pi$ in the time-like
region as input. The other
one is to calculate the correlation function using the operator product
expansion (OPE):
\begin{eqnarray}
  \label{2.13}
  \frac{12 \pi}{Q^2} \Pi(q^2=-Q^2)& =&
  \frac{d}{ \pi} \left[-(1+\alpha_S(Q^2)/\pi) 
  \ln{\left(\frac{Q^2}{\mu^2}\right)} \right. \\ \nonumber && +\left. \frac{c_1}{Q^2}+\frac{c_2}{Q^4}+\frac{c_3}{Q^6}+... \right].
\end{eqnarray}
Here $d$ agrees with the value of the perturbative continuum (e.g. $d$=3/2 for the
isovector channel) and the coefficients $c_{1,2,3}$ incorporate the non-perturbative parts coming
from the condensates such as the gluon condensate in $c_2$ or the four-quark
condensate in $c_3$. In medium these coefficients become
density dependent because the condensates change in matter and new condensates arise. We use the values
$c_2^{\rho,\omega}=0.04+0.018(\rho/\rho_0) \, \GeV^4$ and $
c^{\rho,\omega}_3=-0.07+0.036 (\rho/\rho_0) \GeV^6$ proposed by Hatsuda et al.
\cite{9} for the $\rho$ and $\omega$ meson. They neglected higher twist
condensates which are hardly known but might be important. For the four-quark
condensates we assume as in ref.\cite{9} that ground state saturation holds to
the same extent as in the vacuum. The coefficient $c_1$ is proportional to
the squared quark mass; due to the small mass of the up and down quark we can
safely neglect those contributions and set $c_1^{\rho,\omega}=0$. The situation
is different in case of the $\phi$ meson. Because of the large strange quark
mass $c_1$ is no longer negligible and we take the values
$c^\phi_1=-0.07\, \GeV^2$, $c_2^{\phi}=-0.1+0.01 (\rho/\rho_0) \,
\GeV^4$ and $c^\phi_3=-0.07+0.006 (\rho/\rho_0) \GeV^6$.

A Borel transform is used in order to improve the convergence of the OPE
series. Comparing eq.(\ref{2.12}) and (\ref{2.13}) after Borel transformation we end up with
\begin{eqnarray}
  \label{2.14} 
  \frac{12 \pi^2
  \Pi(0)}{d \M^2}+\frac{1}{
  d \M^2} \int_0^{\infty}ds R (s)
  \exp{\left[-\frac{s}{\M^2}\right]}= \nonumber \\
 \hspace{2cm}  (1+\alpha_S(\M^2)/\pi)+\frac{c_1}{\M^2}+\frac{c_2}{\M^4}+\frac{c_3}{2 \M^6},
\end{eqnarray}
where the Borel mass parameter should be chosen in the range $\M \gsim 1 \,
\GeV$ in order to ensure convergence of the OPE side of eq.(\ref{2.14}).  The
``left side'' comes from the dispersion relation of eq.(\ref{2.12}) and $R$
represents the ratio (\ref{2.6}), but now specified for each individual flavour
channel with $V=\rho,\omega,\phi$. The ``right side'' is determined by the OPE.
In fig. 3a, b and c we show the comparison between the ``left side'' (solid
line) and the ``right side'' (dashed line) for the various channels.
The consistency between ``left'' and ``right'' sides in the vacuum is evidently quite
satisfactory. Only for a Borel mass below $0.8 \, \GeV$ higher order terms in
the OPE become important and the convergence fails. 
The agreement at normal nuclear matter density is not as excellent as in the
vacuum but still very impressive.  We therefore conclude that QCD sum rules
and our hadronic model of the in-medium current-current correlation
function are mutually compatible. We also point out that assuming a simple pole ansatz for the spectrum at finite densities can lead
to erroneous interpretation of the in-medium masses. This is very obvious in the isovector
channel. While our model gives only a small change of the rho mass a simple
pole ansatz leads to a reduction of $m_\rho$ by more than 10 percent \cite{9} at $\rho=\rho_0$.
\begin{figure}[h]
\unitlength0.7071mm
\begin{picture}(208,180)
\put(50,90){\makebox{\epsfig{file=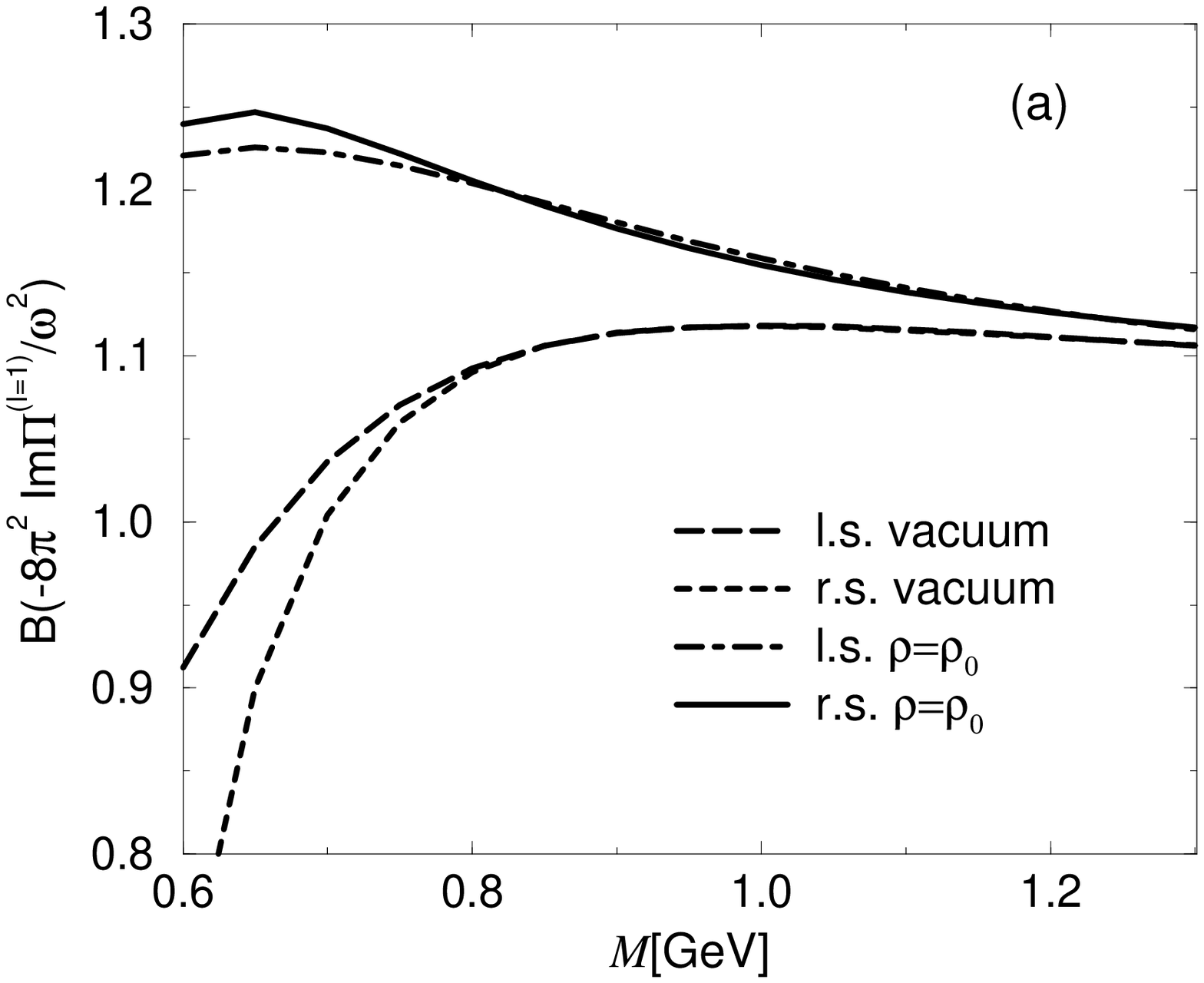,width=75mm}}}
\put(0,0){\makebox{\epsfig{file=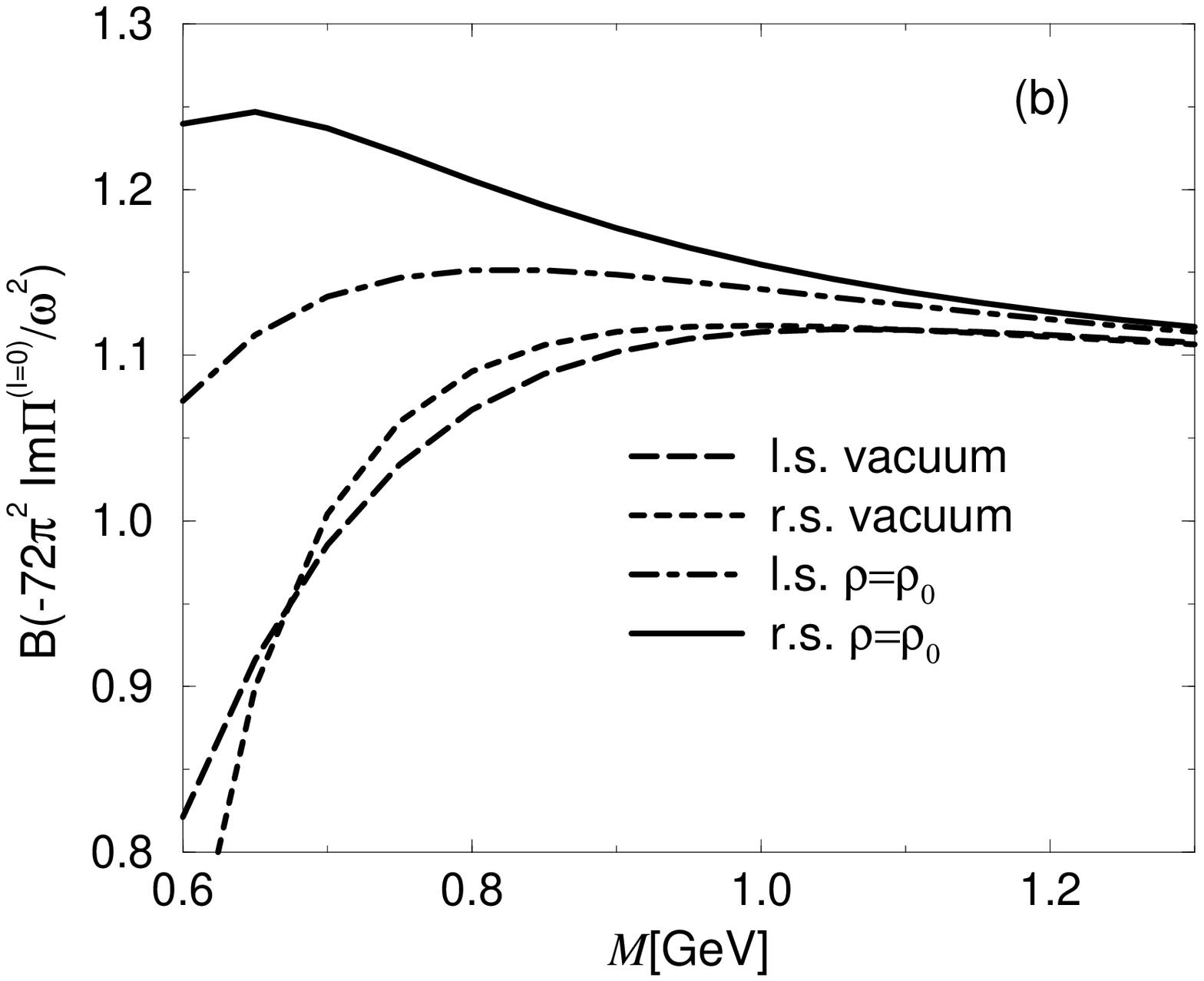,width=75mm}}}
\put(100,0){\makebox{\epsfig{file=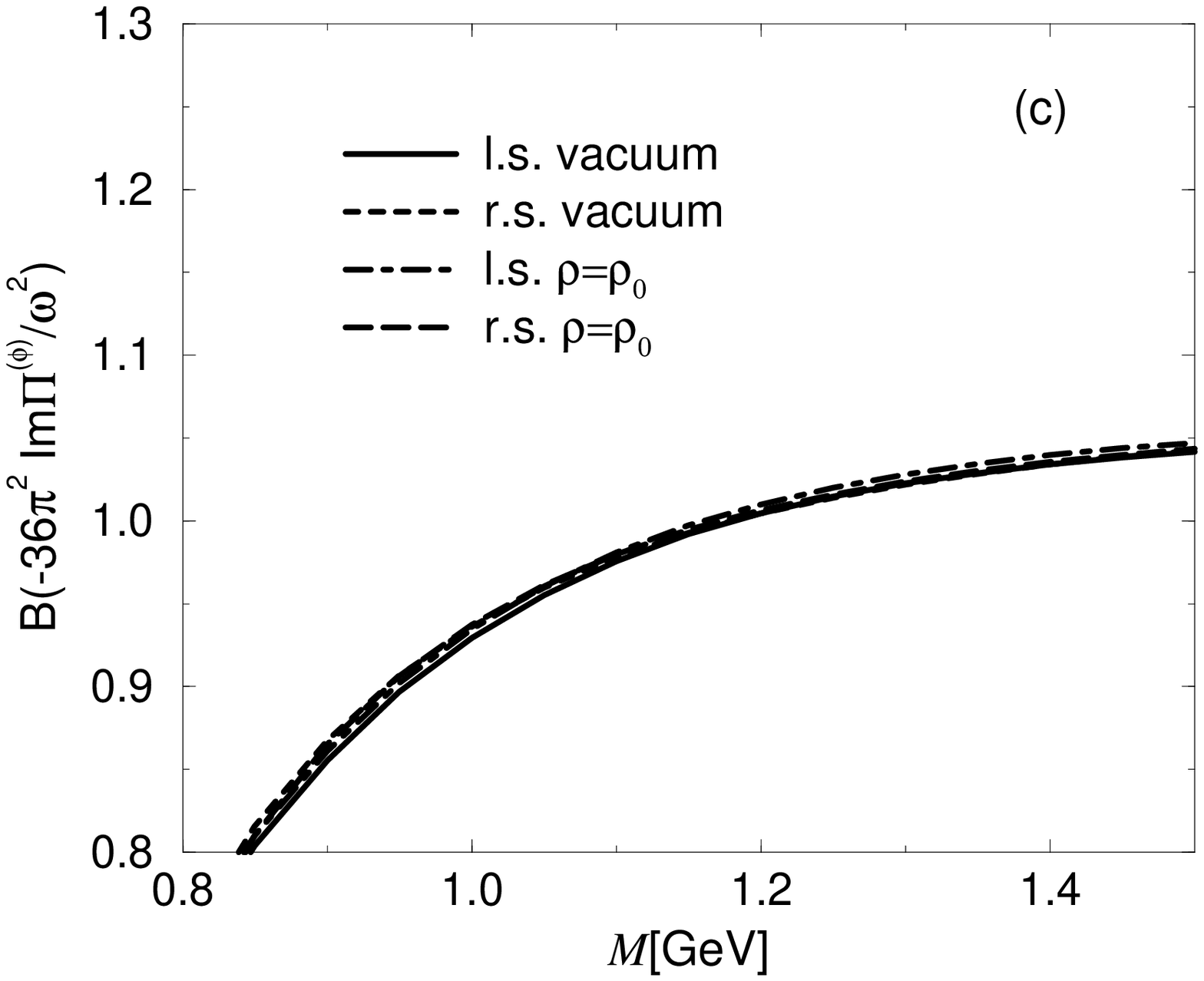,width=75mm}}}

\put(95,175){\makebox{a: $\rho$ meson}}
\put(45,85){\makebox{b: $\omega$ meson}}
\put(145,85){\makebox{c: $\phi$ meson}}
\end{picture}
\caption{Comparison of the ``left'' and ``right'' side of the QCD sum rules
(15) in the
vacuum (solid and dashed line) and at
normal nuclear density (dot dashed and long dashed line) as a function of the
Borel mass parameter { \protect \cal M}.}
\vspace*{-2cm}
\end{figure}
\section{Can one observe chiral restoration?}

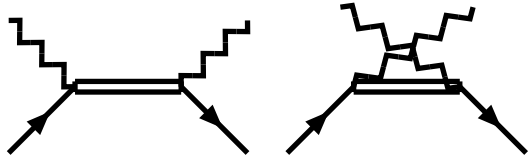
\begin{figure}
\begin{center}
\begin{picture}(200,60)(20,20)
\SetWidth{2} 
\ArrowLine(5,5)(30,30)
\ZigZag(5,55)(30,30){3}{3}
\Line(30,28)(70,28)
\Line(30,32)(70,32)
\Line(30,28)(30,32)
\Line(70,28)(70,32)
\ArrowLine(70,30)(95,5)
\ZigZag(70,30)(95,55){3}{3}

\ArrowLine(110,5)(135,30)
\ZigZag(130,60)(175,30){3}{4}
\Line(135,28)(175,28)
\Line(135,32)(175,32)
\Line(135,28)(135,32)
\Line(175,28)(175,32)
\ArrowLine(175,30)(200,5)
\ZigZag(135,30)(180,60){3}{4}
\end{picture}
\end{center}
\caption{\label{diag6} The Isobar model for the p-wave interaction of the
$\rho$-meson (wavy-line) with the nucleons (solid line). The double line are
the resonances, where explicitly include the $\Delta(1232),\, N^*(1720)$ and
$\Delta^*(1905)$.}
\end{figure}

\begin{figure}
\vspace*{-1cm}
\begin{minipage}{14cm}
\begin{center}
\unitlength0.85mm
\begin{picture}(208,100)
\put(0,0){\makebox{\epsfig{file=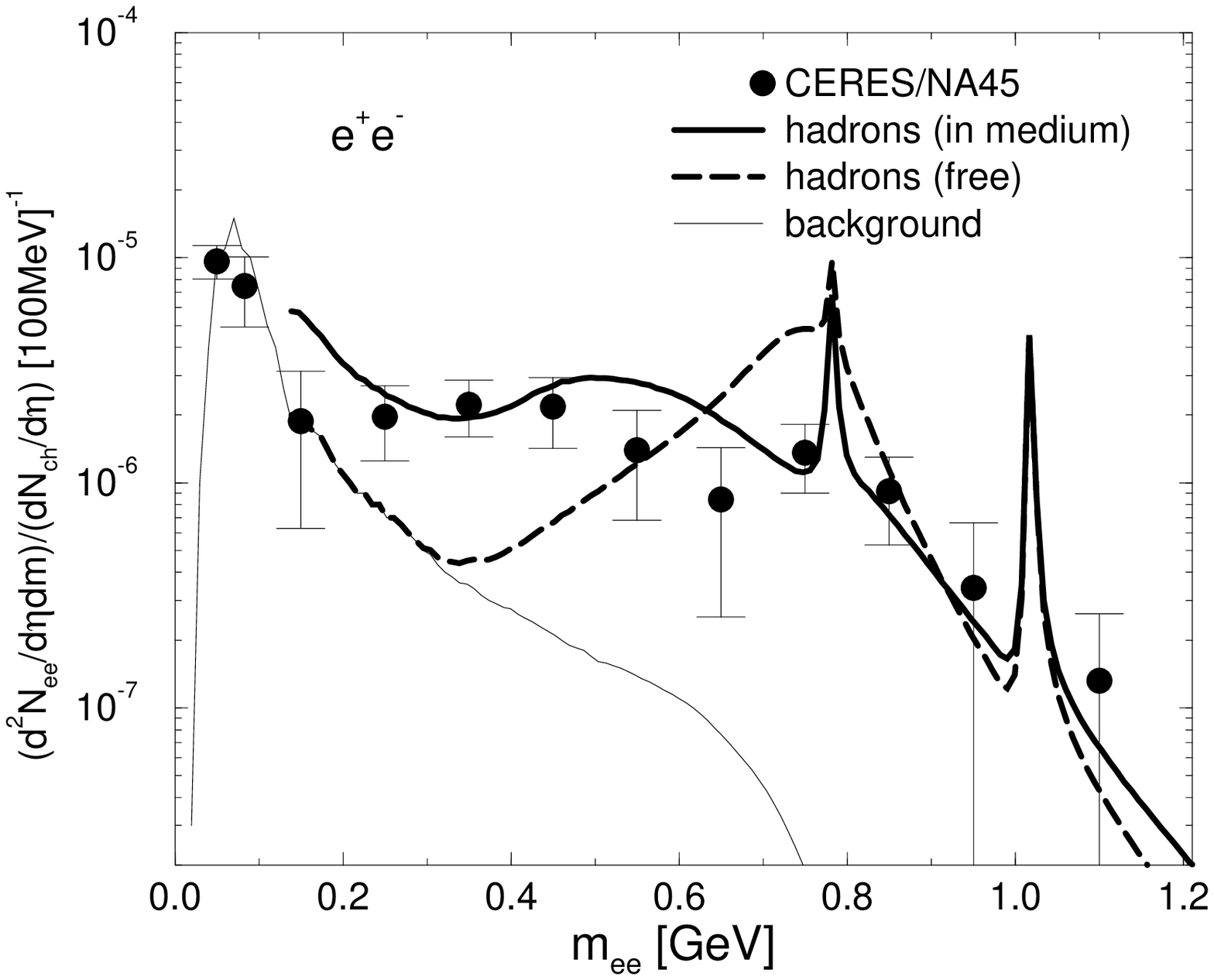,width=79mm}}}
\put(71,0){\makebox{\epsfig{file=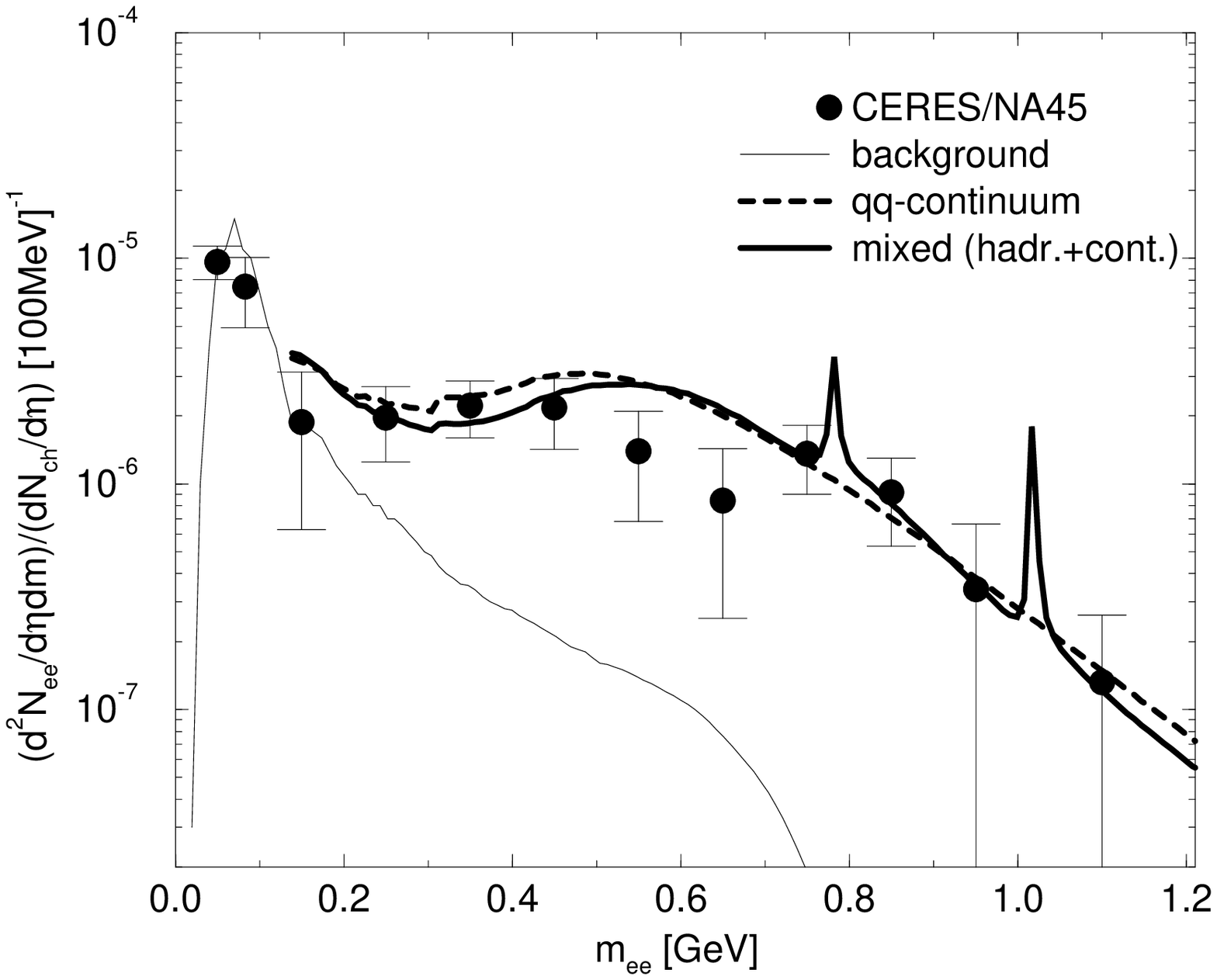,width=79mm}}}
\put(20,20){\makebox{a)}}
\put(91,20){\makebox{b)}}
\end{picture}
\end{center}
\end{minipage}
\vspace*{-0.5cm}
\caption{\label{fig12} Comparison of the dilepton rates from sulfur on gold
collisions measured by the  CERES collaboration \protect \cite{5} with
\protect \newline a)a hadronic system assuming medium or vacuum properties 
\protect \newline b)a system of uncorrelated quark and gluons or a mixed
system of this and a hadronic phase.}
\end{figure}

We now want to compare our result of the correlation function with the dilepton
data measured by CERES \cite{5}. Since the dileptons at this high energy
collisions carry momenta which are of the same order as their invariant mass, we
also have to include the p-wave modification of the correlation
function. Friman and Pirner \cite{14} used an isobar model shown in fig. \ref{diag6} in
order to take the p-wave interaction of the $\rho$-meson with the nucleons into
account. They include in particular the $N^*(1720)$ and the $\Delta^*(1905)$
resonance. For the other vector mesons no such resonances are known. This gives rise
to a momentum dependent vector meson-nucleon scattering amplitude $T_{VN} (
  \omega,\vec{q})$ which we also include in eq.(\ref{3.5}). 
Under the assumption that the change of the spectrum is mainly due to the
density, we can use the resulting $\Im \, \Pi (q,\, \rho,\, T=0) $ as input for eq.(\ref{eq1}) and
determine the thermal dilepton rates. Integrating this over the different
momenta of the dileptons, the volume $V(t)=N_B/\rho(t)$ and the time evolution of the
fireball, we are able to estimate the rate of dilepton pairs emitted in a heavy ion
collision of sulfur on gold at CERN:
\begin{equation}
  \label{ce4}
  \frac{d^2N_{e^+ e^-}}{d\eta\, d m_{e^+ e^-}}= \int_0^{t_f}
  dt V(t) \int d^3 q  \frac{m_{e^+ e^-}}{q_0} \cdot
  \frac{dR(q,\rho(t),T(t))}{d^4x d^4q} A(q).
\end{equation}
where $A(q)$ is the acceptance of the CERES detector. Now we just have to
assume a suitable density and temperature profile which we adapt from the usual
transport code calculations \cite{19}:
\begin{equation}
  \label{ce1}
  T(t)=(T^i-T^\infty)e^{-t/\tau_1}+T^\infty
\end{equation}
with an initial temperature $T^i=170 \, \MeV$. For the final temperature we
take 
$T^\infty=110 \, \MeV$ and we use $\tau_1=8\, \fm$. For the density profile we use
\begin{equation}
  \label{ce2}
  \rho(t)= \rho_i \, exp(-t/\tau_2);      
\end{equation}
with $\rho_i=2.5 \rho_0$ and  $\tau_2=5 \, \fm$. We show the result in
fig.\ref{fig12}a and compare it with the case where we use the free
spectrum at $\rho=0$. One clearly observes a strong improvement as compared to the free
case. 

Now we want to study whether chiral restoration might already occur in such
collisions and assume, as an extreme limit, as system of free, uncorrelated
quarks and gluons. Then the
spectrum of the correlation function would have the constant value of the
perturbative plateau in eq.(\ref{2.10}). We therefore assume $-12 \pi/q^2 \,\Im
\Pi(q^2) =5/3$ below the strange quark threshold and 2 above. The result is
shown in fig.\ref{fig12}b using otherwise the same temperature and baryon density profile as in
the hadronic case. The consistency with the data is astonishing. It is clear,
however, 
that in a realistic scenario both hadronic and quark-gluon phases must be
present. If the initial fireball temperature is as large as $T^i=170 \MeV$, the
system should have already a substantial quark-gluon component. We therefore
assume next that the uncorrelated quarks exist at temperatures above 140
MeV. Below that temperature, the hadronic scenario sets in. The result is also
shown in fig.\ref{fig12}b. 

Considering roughly comparable magnitudes of dileptons produced in both, the hadronic scenario and the phase of
uncorrelated quark and gluons, the real problem is
how one can distinguish them. It is clear that at the moment the precision of
the CERES data opens no way to separate them. A possible signal might be the
suppression of the $\omega$- and $\phi$-meson peak in the quark-gluon phase. But
also a strong broadening of these resonances in the hadronic phase can lead to
a similar suppression. It is therefore essential first to study the hadronic
properties more precisely, and under more moderate conditions. 

\begin{figure}[ht]
\vspace*{-1 cm}
\begin{minipage}{14cm}
\begin{center}
\unitlength1mm
\begin{picture}(70,75)
\put(0,0){\makebox{\epsfig{file=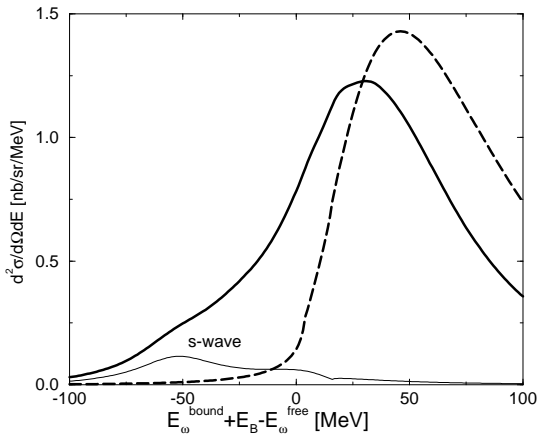,width=80mm}}}
\end{picture}
\end{center}
\end{minipage}
\vspace*{-0.5cm}
\caption{\label{fig14} Typical spectrum of the transfer reaction $d+A\to
^3\!He+(A-1)+\omega$. We consider the results from the free potential (dashed
line) and from our in-medium potential (solid line)} 
\end{figure}

One way is to look for bound $\omega$-meson states. In our hadronic model for
example we observe a strong attractive potential for the $\omega$-meson in
nuclear matter. This behaviour is also supported by QCD sum rules. Such an
attractive potential would lead to bound $\omega$ meson states already in a
light nucleus like Lithium. Such states could be produced via a transfer reaction
like $d+A \to ^3\!He+(A-1)+\omega$ \cite{20}. Here the incoming deuteron $d$ picks
up one proton from the target nucleus $A$ and is detected as $^3\!He$ in forward
direction. During this process the $\omega$-meson is produced and and possibly
bound in the residual $(A-1)$-nucleus. The preferred deuteron beam energy is
such that the $\omega$ is produced with minimal possible momentum. The momentum and energy conservation determines then the energy of
the bound $\omega$-meson. A typical spectrum which one expects on the basis of
a realistic distorted wave Green's functions calculation, is shown in
Fig.\ref{fig14}.
We recognize that the increased medium width broadens the single bound $\omega$
strongly and makes it difficult to identify separate levels. Nevertheless, one
would expect to observe a strong downward shift of the spectral weight which
would be a signal for an attractive medium potential. Of course, the
differential cross section for this process is small, and a detailed
examination of competing background (not taken into account in Fig. 8) is
necessary.

\section{Conclusions}
We have compared the dilepton rates arising from a hadronic model with those coming
from a phase of uncorrelated quarks. Both scenarios describe the
CERES-data well and can not be distinguished at the present level of accuracy.
More precise data under less extreme thermodynamic conditions are needed,
e.g. from the forthcoming HADES experiments at GSI which permit better $e^+e^-$
resolution.  One should
also study the possibility of bound states of $\omega$-mesons in order to gain more information and
better understanding.

\end{document}